\documentclass[twocolumn,notitlepage,prb,superscriptaddress,longbibliography,nofootinbib]{revtex4-2}
\usepackage{amsfonts}
\usepackage{textcomp}
\usepackage{times}
\usepackage{graphicx}
\usepackage{float}
\usepackage{latexsym,amsmath,amssymb,bm,euscript}
\usepackage{xcolor}
\usepackage{subfigure}
\usepackage{epstopdf}
\usepackage[colorlinks=true,linkcolor=blue,citecolor=blue,urlcolor=blue]{hyperref}
\usepackage{hyperref}
\usepackage{soul}
\usepackage[normalem]{ulem}
\usepackage{mathrsfs}
\usepackage{amsmath}
\usepackage{amstext}
\usepackage[bottom]{footmisc}
\usepackage{amsbsy}
\usepackage{ulem}

\begin{document}

\title{Tuning competition between charge order and superconductivity in the square-lattice $t$-$t'$-$J$ model}

\author{Xin Lu}
\affiliation{School of Physics, Beihang University, Beijing 100191, China}

\author{Huaiming Guo}
\email{hmguo@buaa.edu.cn}
\affiliation{School of Physics, Beihang University, Beijing 100191, China}

\author{Wei-Qiang Chen} 
\email{chenwq@sustech.edu.cn} 
\affiliation{Department of Physics and Guangdong Basic Research Center of Excellence for Quantum Science, Southern University of Science and Technology, Shenzhen 518055, China}
\affiliation{Shenzhen Key Laboratory of Advanced Quantum Functional Materials and Devices, Southern University of Science and Technology, Shenzhen 518055, China}

\author{D. N. Sheng}
\email{donna.sheng1@csun.edu}
\affiliation{Department of Physics and Astronomy, California State University Northridge, Northridge, California 91330, USA}

\author{Shou-Shu Gong}
\email{shoushu.gong@gbu.edu.cn}
\affiliation{School of Physical Sciences, Great Bay University, Dongguan 523000, China, and \\
Great Bay Institute for Advanced Study, Dongguan 523000, China}

\date{\today}
\begin{abstract}
\end{abstract}

\begin{abstract}
Recently, a flurry of works have found strong competition between charge density wave (CDW) and superconductivity (SC) in the doped Hubbard and $t$-$J$ models on the square lattice.
%
Interestingly, some recent results suggest that the electron-phonon coupling may suppress CDW order and enhance SC.
In this work, we consider the square-lattice Hubbard model with the Holstein or Su-Schrieffer-Heeger electron-phonon coupling at the large-$U$ and antiadiabatic (infinite phonon frequency) limit, which gives an effective $t$-$J$ model with either a density attractive interaction $V$ or a $J_P$ term that contributes a larger spin exchange and a density repulsive interaction.
To explore how these effective couplings may tune the competition between CDW and SC, we implement the density matrix renormalization group simulations on the $t$-$t'$-$J$ model with $V$ or $J_P$ coupling.
We focus on the {\it six-leg} cylinder system with the next-nearest-neighbor hopping $t'$, which hosts partially filled stripe and $d$-wave SC in phase diagram. 
By tuning $t'/t > 0$ and $V$ or $J_P$, we establish two quantum phase diagrams at $1/12$ doping ratio.
In the SC phases, the increased $V$ or $J_P$ coupling can enhance the quasi-long-range SC order, consistent with some previous findings.
Nonetheless, no SC emerges when the partially filled stripe phase disappears with increased $V$ or $J_P$.
Instead, the system has a transition to either a phase-separation-like regime or a filled stripe phase. 
On the other hand, with increased $t'/t$, not only the partially filled stripe but the phase separation and filled stripe can also be tuned to SC phase.
Our results suggest that although $V$ and $J_P$ couplings may strengthen hole binding, the hole dynamics controlled by $t'/t$ appears to play more crucial role for obtaining a SC in $t$-$J$ model.
\end{abstract}

\maketitle

\section{Introduction}
Understanding the unconventional superconductivity (SC) in cuprate superconductors is a central theme in condensed matter physics~\cite{Keimer_nature_2015,Proust_ARCMP_2019}.
Because the cuprate SC is usually realized by doping the parent Mott insulators, the large-$U$ Hubbard model and the related $t$-$J$ model with no-double-occupancy constraint have been widely considered as the minimal systems to study the emergent SC in doped Mott insulators~\cite{Anderson_Science_1987,Lee_RMP_2006,Arovas_ARCMP_2022,Anderson_2004,Masao_RPP_2008,Zhang_three_band}. 
After extensive studies for decades, while analytical solutions for two-dimensional (2D) correlated systems may be still less controlled, recent numerical simulations have made significant progress in the identification of quantum phases~\cite{White_PRL_1998,White_PRB_1999,Sorella_PRL_2002,Himeda_PRL_2002,White_PRL_2003,Corboz_PRL_2014,Simons_PRX_2015,Ehlers_PRB_2017,Garnet_Science_2017,EWHuang_Science_2017,Ido_PRB_2018,Jiang_PRB_2018,Ponsioen_PRB_2019,Qin_PRX_2020,Xu_PRR_2022,Shih_PRL_2004,Raczkowski2007,Capello2008,White_PRB_2009,Corboz_PRB_2011,Dodaro_PRB_2017,Jiang_Science_2019,Chung_PRB_2020,Jiang_PRR_2020,Wietek_PRX_2021,White_PNAS_2021,Jiang_PRL_2021,Gong_PRL_2021,Jiang_PNAS_2022,Wietek_PRL_2022,White_PRB_2022,Qu2022,Jiang_Kivelson_Lee_2023,zhang2023frustrationinduced,Zhang_CPL_2023,lu2023emergent,tJ_Feng_2023,Jiang2023t,xu2023coexistence,Shen_2024}.
Earlier simulations by combining various methods found consistent evidence that the square-lattice large-$U$ Hubbard model near the optimal doping has a stripe order~\cite{Garnet_Science_2017,EWHuang_Science_2017,Qin_PRX_2020,Xu_PRR_2022}, which is characterized by the intertwined charge density wave (CDW) and spin density wave (SDW), but without SC order.
In the region that the excess holes concentrate, the magnetic order is weakened and there is a $\pi$-phase shift of the antiferromagnetic order, which agrees with the description of the Hartree-Fock calculation~\cite{Stripe_MF_Zaanen,Stripe_MF_Schulz,Stripe_MF_Poilblanc,Machida_1,Machida_2}.

The observed stripe state leads to the consideration of other ingredients in the systems with the purpose of suppressing stripe order and enhancing SC. 
For cuprate superconductors, the ARPES measurements strongly suggest the existence of the next-nearest-neighbor (NNN) hopping $t'$~\cite{Kim_PRL_1998,Pavarini_PRL_2001,Damascelli_RMP_2003,Tanaka_PRB_2004}.
By comparing theoretical and experimental results, $t' > 0$ and $t' < 0$ in the Hubbard and $t$-$J$ models were found proper to describe the electron- and hole-doped cuprates, respectively~\cite{belinicher_generalized_1996}.
Interestingly, the earlier studies on the two-leg $t$-$t'$-$J$ ladder unveiled that $t' > 0$ may enhance the coherent propagation of hole pair~\cite{Martins_PRB_2001}.
On four-leg ladder, density matrix renormalization group (DMRG) calculations on the Hubbard and $t$-$J$ models have not only identified a filled stripe phase near $t' = 0$, but also found different Luther-Emery liquid phases with coexistent quasi-long-range SC and CDW orders by tuning both $t' > 0$ and $t' < 0$~\cite{Jiang_Science_2019,Jiang_PRR_2020,Chung_PRB_2020}.

Towards wider systems, DMRG studies have identified a robust $d$-wave SC phase with suppressed charge order at a moderate $t' > 0$ near the optimal doping, for both the $t$-$J$ (six and eight legs)~\cite{White_PNAS_2021,Jiang_PRL_2021,Gong_PRL_2021,Jiang_Kivelson_Lee_2023} and Hubbard (six legs) models~\cite{Jiang2023t}.
A SC phase has also been observed in an infinite projected entangled pair states (iPEPS) study of the $t$-$t'$ Hubbard model~\cite{zhang2023frustrationinduced}.
These numerical results strongly suggest a stable $d$-wave SC in the 2D limit for the systems with a moderate $t' > 0$, which may describe the electron-doped cuprates. 
For the hole doping at $t' < 0$, CDW and SC demonstrate a subtler competition.
While many previous studies including DMRG, iPEPS, and variational Monte Carlo found CDW states~\cite{White_PRB_1999,Himeda_PRL_2002,Ponsioen_PRB_2019}, recent DMRG results rekindle hopes of finding SC.
For the six-leg $t$-$J$ model at $t' < 0$, the partially filled stripe phase is robust near the optimal doping, but at the smaller doping level, e.g. from $1/24$ to $1/36$, the stripe phase is replaced by a quasi-long-range SC with much weakened CDW order~\cite{lu2023emergent}.  
On the wider eight legs, both the canonical ensemble (at $1/8$ doping)~\cite{lu2023emergent} and grand canonical ensemble (near $1/8$ doping)~\cite{tJ_Feng_2023} calculations surprisingly find that the CDW order observed in previous DMRG studies tends to be melted by keeping much larger bond dimensions, accompanied by an emergent $d$-wave SC.  
In addition, based on the grand canonical ensemble simulations, there is a possible co-existing SC and pair density wave order in the eight-leg $t$-$t'$-$J$ model around $t'=0$~\cite{tJ_Feng_2023}.
These new DMRG results indicate that for the eight-leg $t$-$t'$-$J$ model at $t' \leq 0$, an extremely high accuracy is required to converge to the true ground state.  
Importantly, with increasing system circumference, SC can win the charge order, which sheds new light on the model study of the hole-doped cuprates.

On the other hand, there is growing evidence that electron-phonon couplings (EPC) are also relevant in some cuprate materials~\cite{Exp_EPC0,Exp_EPC1,Exp_EPC2,Exp_EPC3,Exp_EPC4,Exp_EPC5,Exp_EPC6,Exp_EPC7,Exp_EPC8,eph_shenzhixun_2021}. 
For example, the combined spectroscopic and theoretical studies of one-dimensional cuprate Ba$_{2-x}$Sr$_x$CuO$_{3+\delta}$ have unveiled an anomalously strong nearest-neighbor (NN) attractive interaction~\cite{eph_shenzhixun_2021}, which can be derived from the Holstein EPC in the antiadiabatic limit (large phonon frequency limit)~\cite{eph2_shenzhixun_2021}.
The experimental findings have stimulated broad interests in the additional attractive interaction in Hubbard model~\cite{linhaiqing_2021,jiangmi_2022,matianxing_2022,wanghaoxin_2022_1D,Qu_HubbardV_2022,Peng_2023,changjun_2023,xu2024,eph3_shenzhixun_2023,Cheng_PRB_2024}.
It has been found that in the SC or quasi-long-range SC phases, SC order can be further enhanced by a moderate NN density attractive interaction~\cite{Qu_HubbardV_2022,Peng_2023,changjun_2023,xu2024}.
Meanwhile, EPC has been considered directly in the Hubbard and $t$-$J$ models, such as the Holstein EPC~\cite{wangyao_PRR_2020,eph_Karakuzu_2022,eph_wangyao,Dobry_1995,Clay_PRL_2005,Tam_PRB_2007,Tezuka_PRB_2007} and Su-Schrieffer-Heeger (SSH) EPC~\cite{zhangchao_eph_2023,eph2_lizixiang_2022,eph_wanghaoxin,Sous_PRB_2023,Sous_PRL_2018,Xing_SSH_2023,Tanjaroon_eph_2023}.
In the Hubbard-SSH model, DMRG study on the four-leg system finds that the cooperation of EPC and Hubbard repulsion can suppress CDW order and lead to a Luther-Emery liquid phase with coexisted quasi-long-range SC and CDW orders~\cite{eph_wanghaoxin}.
These findings naturally trigger the interests in the questions including whether a moderate EPC can generally suppress CDW orders and stabilize a SC, and whether the cooperation of $t'$ and EPC can make it easier to obtain a SC phase. 

\begin{figure}[t]
	\includegraphics[width=0.95\linewidth]{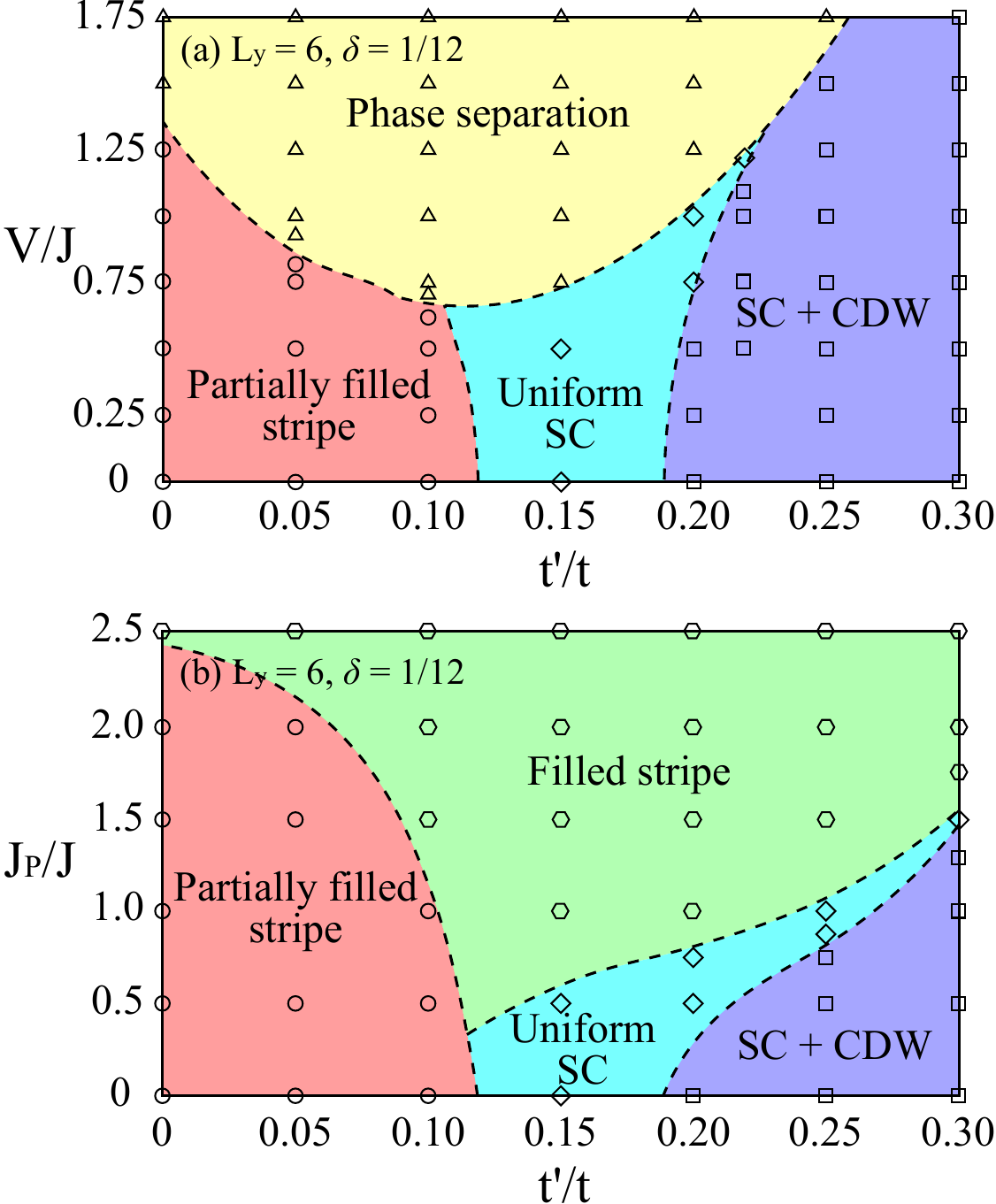}
	\caption{Quantum phase diagrams of the $t$-$t'$-$J$ model with additional interactions on six-leg cylinder at $\delta = 1/12$ doping level. With tuning $0 \leq t'/t \leq 0.3$ and either additional charge density interaction $0 \leq V/J \leq 1.75$ (a) or spin exchange interaction $0 \leq J_P/J \leq 2.5$ (b), the systems have the partially filled stripe, uniform SC, and SC + CDW phases. At larger $V/J$ ($J_P/J$) coupling, the system shows a phase separation (filled stripe phase). The symbols denote the parameter points that we have calculated by DMRG. The dashed lines representing phase boundaries are guides to the eye.}
	\label{Pha_Dia}
\end{figure}

%
Since the direct DMRG simulation of the Hubbard model with EPC is seriously limited by system size, we consider the $t$-$J$ model with additional effective couplings following the large-$U$ and antiadiabatic limit,
which for materials should be more relevant to the cuprates where large-frequency phonons play a major role in the EPC.
To start from a robust CDW order, we choose the {\it six-leg} system with an established partially filled stripe phase.
We study the quantum phases of the square-lattice $t$-$t'$-$J$ model with either a NN density attraction $V$, or a $J_{P}$ coupling that contributes a larger spin exchange and a repulsive density interaction. 
By tuning the NNN hopping $t' > 0$ and $V$ (or $J_P$) at fixed doping level $\delta=1/12$, we identify various quantum phases including the partially filled stripe, uniform SC, SC + CDW phases that have been reported in previous studies~\cite{Gong_PRL_2021}, as well as a filled stripe phase and a phase-separation-like regime. 
In the SC phases, we also find the enhanced SC pairing correlation with increased $V$ or $J_P$. 
Nonetheless, in the presence of $V$ or $J_P$, the partially filled stripe order appears to be robust until a transition to the phase separation or filled stripe, suggesting that the additional density attraction and spin interaction may be not important to stabilize SC in the $t$-$t'$-$J$ model.
On the other hand, the phase diagrams in Fig.~\ref{Pha_Dia} unveil that the increased $t' > 0$ can drive the system from phase separation or filled stripe to SC phase as well, demonstrating its crucial role for the emergence of SC.
Given the existed hole binding in the stripe phases~\cite{lu2023sign,White_PNAS_2021}, our results suggest that even though these additional $V$ and $J_P$ couplings may strengthen binding energy~\cite{Poilblanc_PRB_1993,Rommer_PRB_2000,Zhu_SR_2014,Cheng_PRB_2024}, they appear to be less efficient to enhance phase coherence. 
In the Summary, we compare the results of this effective $t$-$J$ type model with the recent results of the Hubbard-SSH model, which implies the importance of phonon frequency.

The paper is organized as follow. 
In Sec.~\ref{sec:Model}, we introduce the effective Holstein and SSH EPC Hamiltonians in the antiadiabatic and large-$U$ limit, as well as the detail of DMRG calculation. 
In Sec.~\ref{sec:Results}, we present the numerical results, and the last section Sec.~\ref{sec:Summary} is devoted to the summary and discussion.

\section{\label{sec:Model}Model and method}

The general system to consider the EPC in cuprate superconductors could be the Hubbard model with EPC.
The phonons can couple with electrons in different ways.
Here, we consider the Holstein EPC and SSH EPC.
The Holstein EPC describes the coupling between the electron density and the displacement of the lattice, which is defined as~\cite{Holstein_1,Holstein_2}
\begin{equation}\label{eq:Hols}
H^{\mathrm{Holstein}} =\sum_{i,j } g_{ij} {\hat{n} }_{i } ({\hat{a} }_j^{\dagger } +{\hat{a} }_j )+\sum_i \omega {\hat{a} }_i^{\dagger } {\hat{a} }_i,  
\end{equation}
where $\hat{a}_i^{\dagger }$ and $\hat{a}_i$ are the creation and annihilation operators of the Holstein phonon on the site $i$, $\hat{n}_{i}$ is the electron density operator, $\omega$ is the phonon frequency, and $g_{ij}$ denotes the strength of EPC. 
For the SSH EPC, the phonon and the coupling parts are defined as~\cite{Friedel_1970_PRL,Su_1979_PRL}
\begin{equation}
H^{\mathrm{SSH}} =\gamma\sum_{\left\langle ij\right\rangle ,\sigma } ({\hat{b} }_{ij}^{\dagger } +{\hat{b} }_{ij})({\hat{c} }_{i,\sigma }^{\dagger } {\hat{c} }_{j,\sigma } +{\rm H.c.})+\sum_{\left\langle ij\right\rangle } \omega {\hat{b} }_{ij}^{\dagger } {\hat{b} }_{ij},  
\end{equation}
where $\hat{b}_{ij}^{\dagger }$ and $\hat{b}_{ij}$ are the SSH phonon operators defined on the NN bond $\left\langle ij\right\rangle$, $\hat{c}^{\dagger}_{i,\sigma}$ and $\hat{c}_{i,\sigma}$ are the electron creation and annihilation operators with spin $\sigma$ ($\sigma = \uparrow, \downarrow$) on the site $i$, and $\gamma$ denotes the SSH EPC strength.

In this work, we consider the antiadiabatic limit (phonon frequency $\omega \rightarrow \infty$), where the phonon degrees of freedom can be integrated out, resulting in the effective models that can be simulated accurately on larger system size. 
In this limit, the Hubbard-Holstein model can be obtained as~\cite{Sankar_2014,Macridin_2004_PRB}
\begin{equation}\label{eq:HHM}
H = -t \sum_{\left\langle ij\right\rangle,\sigma}(\hat{c}^{\dagger}_{i,\sigma} \hat{c}_{j,\sigma} + {\rm H.c.}) + U \sum_i \hat{n}_{i \uparrow} \hat{n}_{i \downarrow} - V\sum_{\left\langle ij\right\rangle}\hat{n}_i \hat{n}_j,
\end{equation}
where the density-density attractive interaction $V$ originates from the non-local EPC $g_{ij}$ in Eq.~\eqref{eq:Hols}.
Since the longer-range interactions are much weaker in the high-frequency phonon modes, we only consider the NN attraction in Eq.~\eqref{eq:HHM}.
In the same antiadiabatic limit, the SSH Hubbard model  can be expressed as~\cite{eph_wanghaoxin}
\begin{eqnarray}\label{eq:HSSHM}
H &=& -t \sum_{\left\langle ij\right\rangle,\sigma}(\hat{c}^{\dagger}_{i,\sigma} \hat{c}_{j,\sigma} + {\rm H.c.}) + U \sum_i \hat{n}_{i \uparrow} \hat{n}_{i \downarrow} \nonumber \\
&&+ J_{P} \sum_{\left\langle ij\right\rangle}\big[\hat{{\bf S}}_i \cdot \hat{{\bf S}}_j - \frac{1}{2}(\hat{\Delta}^{\dagger}_i \hat{\Delta}_j + {\rm H.c.}) + \frac{1}{4} \hat{n}_i \hat{n}_j\big],
\end{eqnarray}
where ${\hat{\mathbf{S}} }_i =\frac{1}{2}\sum_{\alpha ,\alpha^{\prime } } {\hat{c} }_{i,\alpha }^{\dagger } \vec{\sigma}_{\alpha \alpha^{\prime } } {\hat{c} }_{i,\alpha^{\prime } }$ is the spin-$1/2$ operator, and ${\hat{\Delta} }_i^{\dagger } ={\hat{c} }_{i,\uparrow }^{\dagger} {\hat{c}}_{i,\downarrow }^{\dagger }$ is the on-site pair creation operator.

We consider a large Hubbard repulsive interaction $U/t = 12$ and the parameter regime with a small ratio of either $V/U$ or $J_P / U$ (see the parameter range in Fig.~\ref{Pha_Dia}).
In such cases, the Mott physics should persist, and we apply the Gutzwiller projection to derive the effective $t$-$J$ type models with the no-double-occupancy constraint. 
The models Eq.~\eqref{eq:HHM} and Eq.~\eqref{eq:HSSHM} can be simplified as~\cite{Sankar_2014,Krishanu_PLA_2019}
\begin{eqnarray}\label{HtJ1}
H_{V}&=& -t\sum_{ \langle ij\rangle,\sigma} (\hat{c}^{\dagger}_{i,\sigma} \hat{c}_{j,\sigma} + {\rm H.c.}) - t^{\prime}\sum_{\langle\langle ij\rangle\rangle,\sigma} ( \hat{c}^{\dagger}_{i,\sigma} \hat{c}_{j,\sigma} + {\rm H.c.}) \nonumber \\
&+& J\sum_{\langle ij\rangle } {\hat{\mathbf{S}} }_i \cdot {\hat{\mathbf{S}} }_j - (\frac{J}{4}+V)\sum_{\left\langle ij\right\rangle } {\hat{n} }_i {\hat{n} }_j,
\end{eqnarray}
and~\cite{Gunnarsson_2004_1,Gunnarsson_2004_2} 
\begin{eqnarray}\label{HtJ2}
H_{J_P}&=& -t\sum_{ \langle ij\rangle,\sigma} ( \hat{c}^{\dagger}_{i,\sigma} \hat{c}_{j,\sigma} + {\rm H.c.} ) - t^{\prime}\sum_{\langle\langle ij\rangle\rangle,\sigma} ( \hat{c}^{\dagger}_{i,\sigma} \hat{c}_{j,\sigma} + {\rm H.c.}) \nonumber \\
&+& (J+J_P )\sum_{\langle ij\rangle } {\hat{\mathbf{S}} }_i \cdot {\hat{\mathbf{S}} }_j-\frac{1}{4}(J-J_P )\sum_{\left\langle ij\right\rangle } {\hat{n} }_i {\hat{n} }_j,
\end{eqnarray}
respectively. 
While the $V$ term enhances the density attraction in Eq.~\eqref{HtJ1}, the $J_P$ coupling in Eq.~\eqref{HtJ2} contributes a larger spin exchange and a repulsive density interaction.
In our study, we also tune the ratio of $t'/t$.
We choose $J=1.0$ as the energy unit and set $t/J= 3.0$ to mimic a strong Hubbard repulsion $U/t=12$~\cite{Misumi_PRB_2017}.
We use DMRG~\cite{White_PRL_1992} to simulate the ground state of the systems.
The length and width of the lattice are denoted as $L_x$ and $L_y$, giving the total site number $N = L_x \times L_y$.
The doping ratio $\delta$ is defined as $\delta = N_h / N$, where $N_h$ is the number of doped holes.
We consider the cylinder geometry with the open and periodic boundary conditions along the axial ($x$) and circumference ($y$) directions, respectively.
We study the six-leg ($L_y = 6$) system and focus on the parameter regime $0\le t'/t \le 0.3$ at $\delta=1/12$. 
In most cases of simulations, we choose $L_x = 48$, which can match the CDW wave vectors of the different phases and is long enough to accurately analyze the decay behavior of correlation functions in real space. We have also carefully examined different system lengths for cross-check, such as $L_{x} = 24, 36, 48, 56, 64$, which give consistent results.
We implement the spin rotational ${\rm SU(2)}$~\cite{I.P.McCulloch_2002} and charge ${\rm U(1)}$ symmetries, and keep the bond dimensions of $\rm SU(2)$ multiplets up to $D = 12000$ in most cases.
For some parameter points, we further increase the bond dimensions to $D = 15000$, which gives accurate results with the truncation error less than $1.7 \times 10^{-6}$. 
In our DMRG calculations for these two systems, we find the bond dimension $D = 6000$ can obtain the qualitatively correct ground state, thus for more DMRG testing shown in the Appendix we use $D = 6000$.

\section{\label{sec:Results}DMRG results}

\subsection{Quantum phase diagrams}

We map out the phase diagrams of the $t$-$t'$-$J$ model with additional either $V$ [Fig.~\ref{Pha_Dia}(a)] or $J_P$ [Fig.~\ref{Pha_Dia}(b)] coupling, based on the results of charge density profile and different correlation functions.
With very small $V$ or $J_P$ coupling, the system shows the same phases that have been reported in the six-leg $t$-$t'$-$J$ model~\cite{Gong_PRL_2021}, including a uniform SC phase, a partially filled stripe phase with the CDW wavevector ${\bf Q} = (3\pi\delta, 0)$, and a coexistent SC + CDW phase with the CDW wavevector ${\bf Q} = (6\pi\delta, 0)$.
In the uniform SC phase, the charge density distribution is uniform in the bulk of the system, and the pairing correlation is dominant among various correlation functions, with a small power exponent describing the quasi-long-range SC order. 
In the SC + CDW phase, the charge density profile is not uniform but shows a weak oscillation, and the spin correlation exhibits the N\'eel antiferromagnetic behavior.

With increased $V$ or $J_P$ interaction, the system enters either a phase-separation-like regime~\cite{PS_Lin,PS_LHG,PS_ZZ} or a filled stripe phase with the CDW wavevector ${\bf Q} = (2\pi\delta,0)$ and one hole filled in each stripe~\cite{Jiang_PRR_2020,Qin_PRX_2020}. 
In the $t$-$J$ model, the phase separation with a hole-rich and a no-hole phase has been found when the system has a relatively strong $J$ term~\cite{PS_LHG,PS_Lin,PS_ZZ}. 
In our simulation, we find that in the regime with a larger $V$, the charge density distribution shows the hole-rich and no-hole regions [Fig.~\ref{CDW_Profile}(b2)], which is consistent with a phase separation.

In the uniform SC and SC + CDW phases, we also find that the SC pairing correlation could be enhanced with increased $V$ or $J_{P}$ interaction.
However, in the partially filled stripe phase, the additional $V$ or $J_P$ coupling can only slightly weaken the CDW order, and no SC emerges when phase transitions happen. 
The phase diagrams in Fig.~\ref{Pha_Dia} demonstrate the additional $V$ or $J_P$ coupling may not make it easier to obtain a SC with increased $t'$, and a moderate $t' > 0$ appears to be crucial.

\subsection{Charge density profile}

\begin{figure}[t]
	\includegraphics[width=1.0\linewidth]{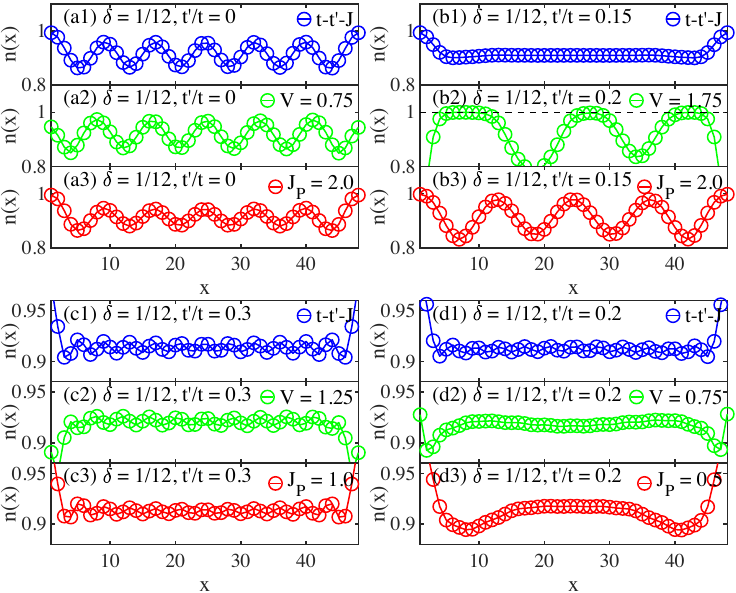}
	\caption{Charge density profile $n(x)$ for different $t'/t$ with additional couplings. (a1-a3) $t'/t = 0$ in the partially filled stripe phase. (b1-b3) $t'/t = 0.15$ or $0.2$ in the uniform SC, phase separation, and filled stripe phase, respectively. (c1-c3) $t'/t = 0.3$ in the SC + CDW phase. (d) $t'/t = 0.2$ in the SC + CDW and uniform SC phases. The results are obtained from $L_x =48$ and by keeping bond dimensions $D=12000$.}
	\label{CDW_Profile}
\end{figure}

We first characterize the different phases by showing the charge density distribution $n(x) = \sum_{y=1}^{L_y} \langle {\hat{n} }_{x,y} \rangle / L_y$, where $x$ denotes the column number of cylinder. 
In the quantum states with a long-range CDW order, $n(x)$ shows a robust density wave oscillation on the cylinder with the open boundary conditions.
If the CDW is a quasi-long-range order, the density wave oscillation will decay algebraically from the boundary to the bulk~\cite{Friedel_oscillations}.

In the partially filled stripe phase, as presented in Fig.~\ref{CDW_Profile}(a1-a3), the CDW pattern of $n(x)$ remains stable in the presence of either a small $V$ or $J_{P}$ coupling [also see Fig.~\ref{CDW_supp} in Appendix A], with the ordering wavevector ${\bf Q} =(3\pi\delta,0)$.
At a moderate $t'/t$, the uniform SC phase [Fig.~\ref{CDW_Profile}(b1)] can extend to the finite-$V$ or $J_{P}$ region.
With further increasing the interactions, the system enters into either a phase-separation-like phase [Fig.~\ref{CDW_Profile}(b2)] or a filled stripe phase with ${\bf Q} = (2\pi\delta,0)$ [Fig.~\ref{CDW_Profile}(b3)]. 

In the SC + CDW phase, we find that the amplitude of charge density oscillation can be gradually suppressed with growing $V$ or $J_{P}$ [see Fig.~\ref{CDW_Profile}(c1-c3) and Fig.~\ref{CDW_supp} in Appendix A], which agrees with the recent observation in the Hubbard model~\cite{xu2024}. 
For some parameter regimes of $t'/t$, the increased $V$ and $J_{P}$ may even melt the bulk charge density oscillation and drive the system to the uniform SC phase, as shown in Fig.~\ref{CDW_Profile}(d1-d3).

\subsection{Correlation functions}

\begin{figure}[t]
	\includegraphics[width=1.0\linewidth]{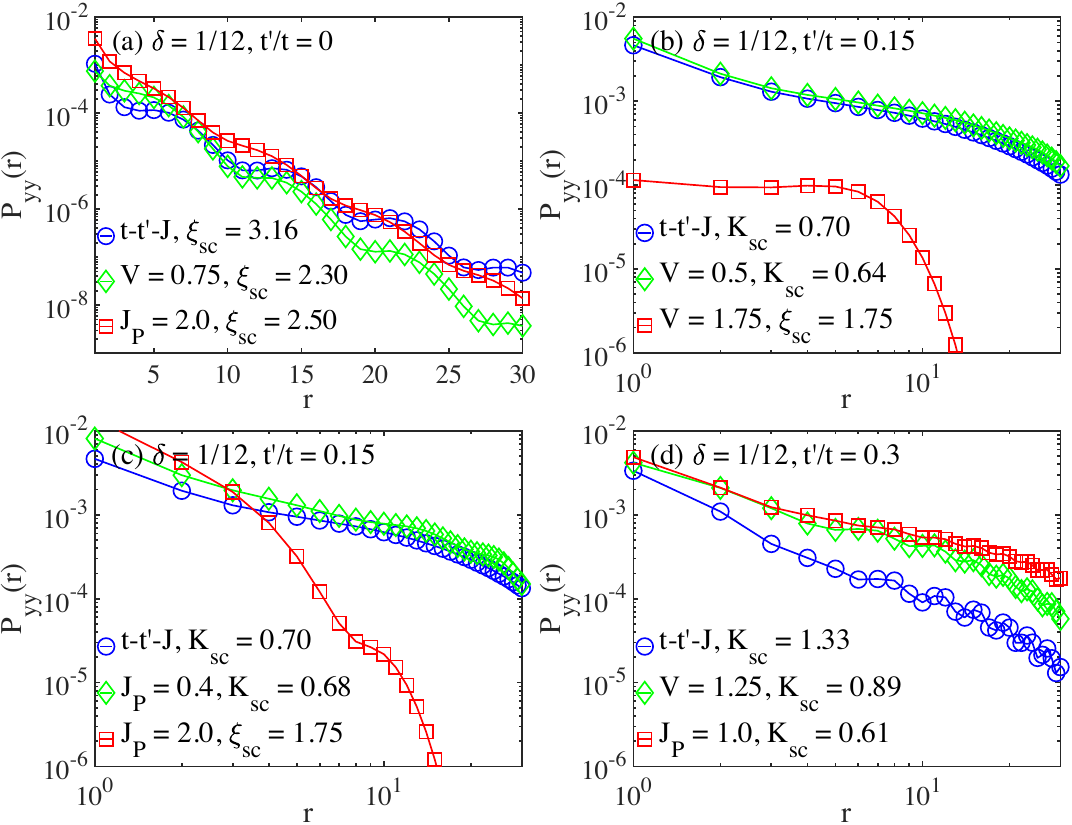}
	\caption{SC pairing correlation function $P_{yy}$ for different $t'/t$ with additional couplings. (a) Semi-logarithmic plot of $P_{yy}$ in the partially filled stripe phase at $t'/t = 0$. (b) Double-logarithmic plot of $P_{yy}$ in the uniform SC phase and phase separation with growing $V/J$ at $t'/t = 0.15$. (c) Double-logarithmic plot of $P_{yy}$ in the uniform SC and filled stripe phases with growing $J_P / J$ at $t'/t = 0.15$. (d) Double-logarithmic plot of $P_{yy}$ in the SC + CDW phase at $t'/t = 0.3$. The correlation length $\xi_{\mathrm{sc}}$ and power exponent $K_{\rm sc}$ are obtained by exponential and power-law fittings, respectively. The data are obtained from $L_x =48$ and by keeping bond dimensions $D=12000$.}
	\label{SC_Pairing_Correlation}
\end{figure}

We further discuss the correlation functions to characterize the different quantum phases.
To examine the behavior of SC order, we calculate the spin-singlet pairing correlations $P_{\alpha, \beta }(\mathbf{r}) = \langle {\hat{\Delta}}_{\alpha }^{\dagger } ({\mathbf{r}}_0) {\hat{\Delta}}_{\beta }({\mathbf{r}}_0 +\mathbf{r}) \rangle$, where the pairing operator is defined as ${\hat{\Delta} }_{\alpha } \left(\mathbf{r}\right)=\left({\hat{c} }_{\mathbf{r}\uparrow } {\hat{c} }_{\mathbf{r}+{\mathbf{e}}_{\alpha } \downarrow } -{\hat{c} }_{\mathbf{r}\downarrow } {\hat{c} }_{\mathbf{r}+{\mathbf{e}}_{\alpha } \uparrow } \right)/\sqrt{2}$ and ${\mathit{\mathbf{e}}}_{\alpha = x,y}$ denote the unit vectors along the $x$ and $y$ directions, respectively.
In the partially filled stripe phase [Fig.~\ref{SC_Pairing_Correlation}(a)], the increased $V$ or $J_{P}$ can barely enhance pairing correlation.
In the uniform SC and SC + CDW phases, pairing correlation could be enhanced with growing $V$ or $J_{P}$, as shown in Fig.~\ref{SC_Pairing_Correlation}(b-d). 
In particular, pairing correlation is enhanced apparently in the SC + CDW phase [Fig.~\ref{SC_Pairing_Correlation}(d) and Fig.~\ref{SM_SC_CDW} in Appendix~\ref{sec:Scaling_SCCDW}].
In the phase-separation-like and filled stripe phases, pairing correlations decay exponentially with a small correlation length.


\begin{figure}[t]
	\includegraphics[width=1.0\linewidth]{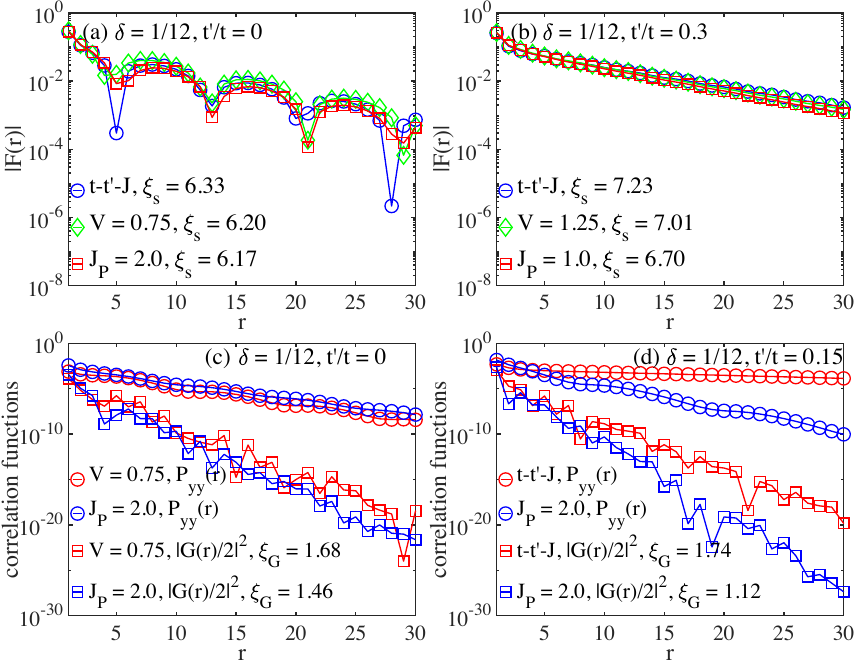}
	\caption{Correlation functions for different models. (a-b) Semi-logarithmic plots of spin correlations $F(r)$ in the partially filled stripe and SC + CDW phases respectively. (c-d) Semi-logarithmic plots of $P_{yy}(r)$ and single electron Green's functions $G^{2}(r)/2$ in the partially filled stripe and SC + CDW phases respectively. The correlation length $\xi_{\mathrm{s}}$ and $\xi_{\mathrm{G}}$ (fitted by $G(r)$) are obtained by exponential fitting. Here, all the data are obtained from $L_x =48$ and by keeping bond dimensions $D=12000$.}
	\label{Spin_Green}
\end{figure}

To describe the magnetic properties, we calculate the spin correlation function $F(r)=\langle  {\hat{\mathbf{S}} }_{x,y} \cdot {\hat{\mathbf{S}} }_{x+r,y}\rangle$. 
In the partially filled stripe phase [Fig.~\ref{Spin_Green}(a)], the increased $V$ or $J_{P}$ barely changes spin correlation, which is consistent with the intertwined robust CDW profile shown in Fig.~\ref{CDW_Profile}(a1-a3). 
In the SC + CDW phase [Fig.~\ref{Spin_Green}(b)], spin correlations always show the N\'eel antiferromagnetic pattern.
We notice that while pairing correlation can be enhanced remarkably with increasing $V$ or $J_P$ in this phase [Fig.~\ref{SC_Pairing_Correlation}(d)], spin correlation is almost invariant [see Fig.~\ref{SM_spin_green}].
In the filled stripe phase, we also find the $\pi$-phase shift of the antiferromagnetic spin correlation, which leads to the period of spin correlation doubled compared with that of the CDW pattern.

Furthermore, we examine the single-particle Green's function $G(r) = \langle \sum_{\sigma} \hat{c}^{\dagger}_{x,y,\sigma}  \hat{c}_{x+r,y,\sigma} \rangle$ in different phases, which always decays exponentially. 
Recent DMRG study of the $t$-$t'$-$J$ model ($L_y = 3, 4, 5, 6$) has unveiled that hole binding exists in both the SC and stripe phases (including the filled stripe at $L_y = 4$ and partially filled stripe at $L_y = 6$)~\cite{lu2023sign}, which implies that the absence of SC in these stripe phases might be owing to the lack of phase coherence.
This feature of the stripe phases appears like the preformed pairs picture of the pseudogap regime, which partially can be characterized by the observation that the pairing correlation decays fast but it is much stronger than two single-particle correlator $G^{2}(r)$~\cite{Huang_PRL_2023_PG}. 
Indeed, in the partially filled stripe phase, even in the presence of either $V$ or $J_P$ coupling, pairing correlation is much stronger than $|G^{2}(r)|$ [Fig.~\ref{Spin_Green}(c)]. 
Interestingly, the similar feature also exists in the filled stripe phase [Fig.~\ref{Spin_Green}(d)], suggesting that the holes may also be paired in this phase.
Since the increased $t'/t$ also drives the transition from the filled stripe to SC phase, one may conjecture that the increased $t'/t > 0$ also enhances the coherent propagation of hole pair in this case, which we leave to future study.

In the Hamiltonian Eq.~\eqref{HtJ2}, $J_P$ contributes both spin exchange and density interactions.
To clarify the consequences from each part, we have also comparatively examined the $t$-$t'$-$J$ model by either increasing the total $J$ term or increasing the Heisenberg part $J \hat{{\bf S}}_i \cdot \hat{{\bf S}}_j$ with the fixed density attraction.
For these model setup, we find similar conclusions to those above, as discussed in Appendix~\ref{sec:Tuning_spin_exchange}.


\section{\label{sec:Summary}Summary and discussion}

We have studied the square-lattice $t$-$t'$-$J$ model with additional couplings in the spin exchange and density-density interaction terms, which may describe the Hubbard model with either the Holstein or the SSH EPC in the antiadiabatic and large-$U$ limit.
Our motivation is to examine whether these additional couplings can suppress the CDW order and result in a SC phase. 
With this purpose, we start from the $t$-$t'$-$J$ model on the {\it six-leg} cylinder, which has the well established partially filled stripe and $d$-wave SC phases near the optimal doping level by tuning the NNN hopping $t' \geq 0$~\cite{Gong_PRL_2021, White_PNAS_2021}.

By means of DMRG calculation, we establish two phase diagrams at $1/12$ doping level [see Fig.~\ref{Pha_Dia}], in both the $t'/t - V/J$ and $t'/t - J_P/J$ planes.
While $V$ denotes the NN density attractive interaction [see Eq.~\eqref{HtJ1}], $J_P$ contributes both a larger spin exchange interaction and a repulsive density interaction [see Eq.~\eqref{HtJ2}].
Our results unveil that in the uniform SC and SC + CDW phases, in which SC has established a quasi-long-range order, pairing correlation can be enhanced by the increased $V$ or $J_P$ coupling.
However, in the partially filled stripe phase the increased $V$ or $J_P$ appears unable to enhance pairing correlation, and no SC phase emerges when this stripe phase disappears.
Instead, the system enters either a phase-separation-like regime or a filled stripe phase at larger $V$ or $J_P$ coupling.
The phase diagrams in Fig.~\ref{Pha_Dia} strongly suggest that the increased $V$ or $J_P$ may be not crucial for obtaining a SC phase in the $t$-$t'$-$J$ model.
For $t' < 0$ on the six-leg system, we find the similar conclusions (see the details in Appendix~\ref{sec:Hole_doped}).
In contrast, our results further support the conception that a moderate $t' > 0$ may play an essential role for stabilizing SC in $t$-$J$ model.

In a recent DMRG study of the six-leg $t$-$t'$-$J$ model, hole binding is found not only in the SC phase but also in the partially filled stripe phase~\cite{lu2023sign}, where the hole pairing is attributed to the phase-string part of an exact sign structure in the partition function, and the sign of $t'$ determines an additional Berry-phase-like term that may account for the different stripe and SC phases. 
Interestingly, our results could be consistent with the inference of this phase-string theory~\cite{Phase_String_PRL_Sheng,Phase_string_PRB_Weng,IJMPB2007}.
While hole pairing already exists in the stripe phase of the $t$-$t'$-$J$ model, the additional $V$ or $J_P$ coupling may strengthen the binding energy but appears not directly connected with the Berry-phase-like term, which can explain our results that SC can emerge with tuning $t'$, but not with increasing $V$ or $J_P$.

As a preliminary study, we only consider the antiadiabatic limit with infinite phonon frequency in this work.
In a recent DMRG study of the four-leg Hubbard-SSH model~\cite{eph_wanghaoxin}, a finite phonon frequency of EPC ($\omega / t = 5$) has been considered, which can drive a transition from stripe to $d$-wave SC near $U/t = 8$. 
However, such a transition is not found in our results at the antiadiabatic limit~\cite{note}, which strongly suggests the importance of phonon frequency and calls for further studies on the role played by phonon frequency in the EPC.

\begin{acknowledgments}
We appreciate the stimulating discussions with Zheng-Yu Weng, Hong Yao, Jia-Xin Zhang, Xun Cai, Hao-Xin Wang, and Junsong Sun. X. L. and S. S. G. were supported by the NSFC (No. 12274014), the Special Project in Key Areas for Universities in Guangdong Province (No. 2023ZDZX3054), the Dongguan Key Laboratory of Artificial Intelligence Design for Advanced Materials, and the Research team on solar metamaterials (2024KCXTD048).
D. N. S. was supported by the US National Science Foundation Grant No. PHY-2216774.
W. Q. C. was supported by the National Key Research and Development Program of China (No. 2024YFA1408101), NSFC (Grants No. 12141402), the Science, Technology and Innovation Commission of Shenzhen Municipality (No. ZDSYS20190902092905285), Guangdong Provincial Quantum Science Strategic Initiative (Grand No. SZZX2401001, GDZX2201001), the SUSTech-NUS Joint Research Program, and Center for Computational Science and Engineering at Southern University of Science and Technology.
H. M. G. was supported by NSFC grant No. 12074022 and the BNLCMP open research fund under Grant No.2024BNLCMPKF023. The computational resources are supported by SongShan Lake HPC Center (SSL-HPC) in Great Bay University.
\end{acknowledgments}

\appendix

\section{Charge density profile}

\begin{figure}[h]
   \includegraphics[width=1.0\linewidth,angle=0]{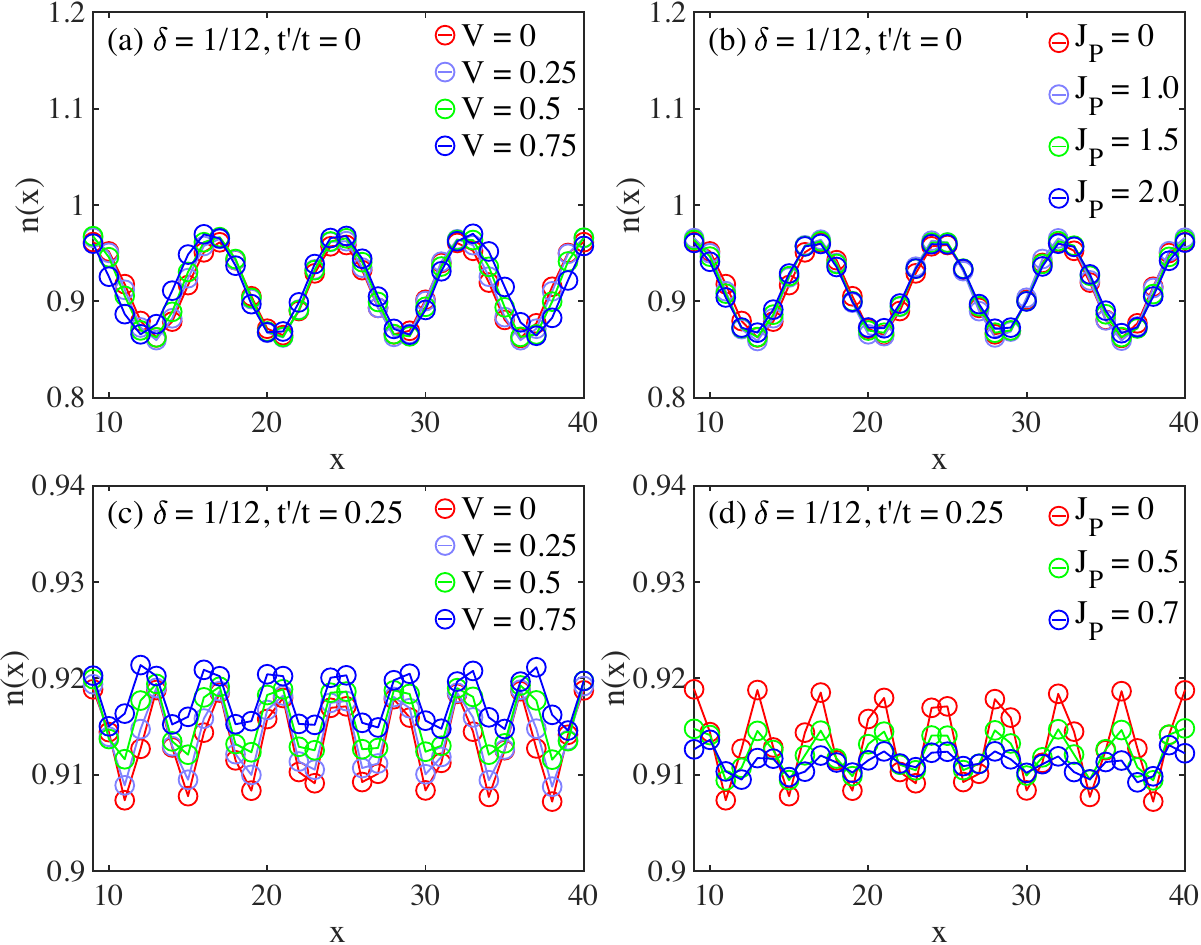}
   \caption{\label{CDW_supp}
   Charge density profile $n(x)$ in the partially filled stripe and SC + CDW phases. (a) and (b) show the results at $\delta = 1/12$, $t'/t = 0$ in the partially filled stripe phase, with additional $V$ and $J_P$ couplings, respectively. (c) and (d) show the similar results at $\delta=1/12$, $t'/t=0.25$ in the SC + CDW phase. Here, all the data are obtained from $L_x =48$ and by keeping bond dimensions $D=6000$. In order to clearly visualize how CDW changes in bulk, a few data points near the boundaries are excluded, and we only show the $n(x)$ within the space $x\in \left\lbrack 9,40\right\rbrack$.
}
\end{figure}

In this section, we show more data of the charge density profile $n(x)$ in the partially filled stripe and SC + CDW phases.
In Figs.~\ref{CDW_supp}(a) and \ref{CDW_supp}(b), we present $n(x)$ in the partially filled stripe phase with increased either $V$ or $J_P$ coupling.
For both cases, charge density profile appears to be stable before the transitions happen.
In the SC + CDW phase, the increased $V$ or $J_P$ can gradually suppress the magnitudes of $n(x)$ as shown in Figs.~\ref{CDW_supp}(c) and \ref{CDW_supp}(d).

\begin{figure}[h]
   \includegraphics[width=1.0\linewidth,angle=0]{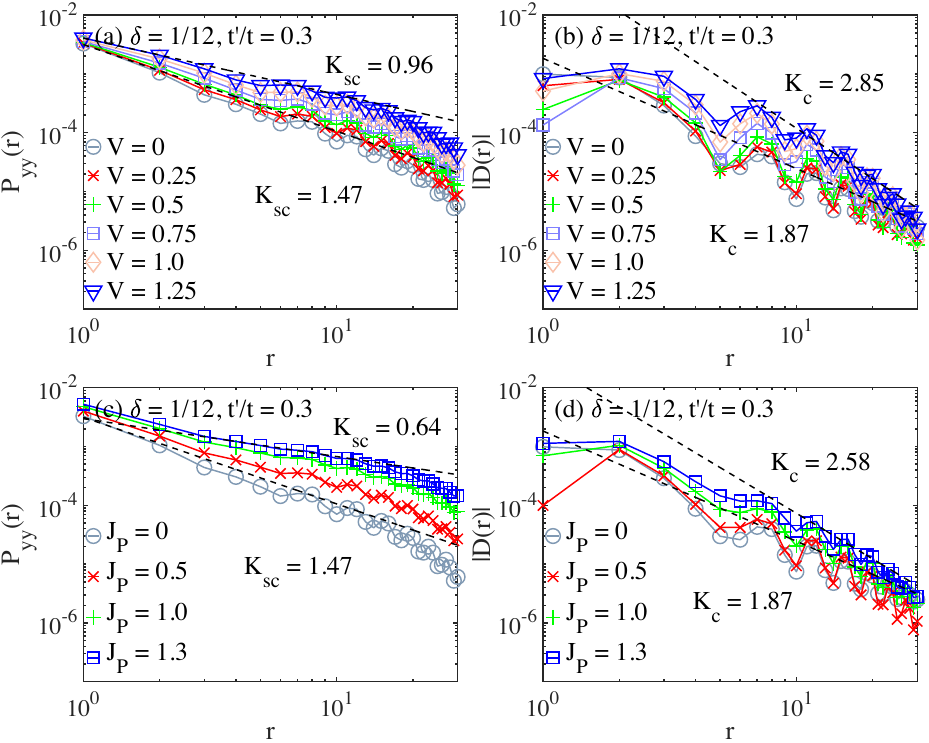}
   \caption{\label{SM_SC_CDW}
   Coupling dependence of the pairing correlation $P_{yy}(r)$ and density correlation $D(r)$ in the SC + CDW phase. (a) and (b) show the results of $P_{yy}(r)$ and $|D(r)|$, respectively, with increased $V$ coupling for $\delta = 1/12$, $t'/t = 0.3$. The power exponents $K_{\mathrm{sc}}$ in (a) are extracted by algebraic fitting of $P_{yy}(r)$ at $V=0$ and $V=1.25$. The exponents $K_{\mathrm{c}}$ in (b) are extracted by algebraic fitting of $|D(r)|$ at $V=0$ and $V=1.25$. (c) and (d) show the similar results with increased $J_P$ coupling. The power exponents $K_{\mathrm{sc}}$ in (c) are extracted by algebraic fitting of $P_{yy}(r)$ at $J_{P}=0$ and $J_{P}=1.3$. The exponents $K_{\mathrm{c}}$ in (d) are extracted by algebraic fitting of $|D(r)|$ at $J_{P}=0$ and $J_{P}=1.3$. All these measurements are obtained from $L_x =48$ and by keeping the bond dimensions $D = 6000$.
}
\end{figure}

\section{\label{sec:Scaling_SCCDW} Correlation functions in the SC + CDW phase}

 
In this section, we present more details of the correlation functions with increased $V$ or $J_P$ coupling in the SC + CDW phase. 
In Fig.~\ref{SM_SC_CDW}, one can find that both the pairing correlation $P_{yy}(r)$ and charge density correlation $|D(r)|$ exhibit a good power-law decay, indicating the quasi-long-range SC and charge orders. 
With increased $V$ or $J_P$ coupling, pairing correlation is enhanced and also decays slower. 
On the other hand, while the magnitude of density correlation is enhanced, it decays faster.
Therefore, $K_{\rm sc}$ and $K_{\rm c}$ have the opposite responses to increased $V$ or $J_P$, which has also been observed in other systems~\cite{Lu_PRB_2023,xu2024,Qu_HubbardV_2022,Peng_2023}.

\begin{figure}[h]
   \includegraphics[width=1.0\linewidth,angle=0]{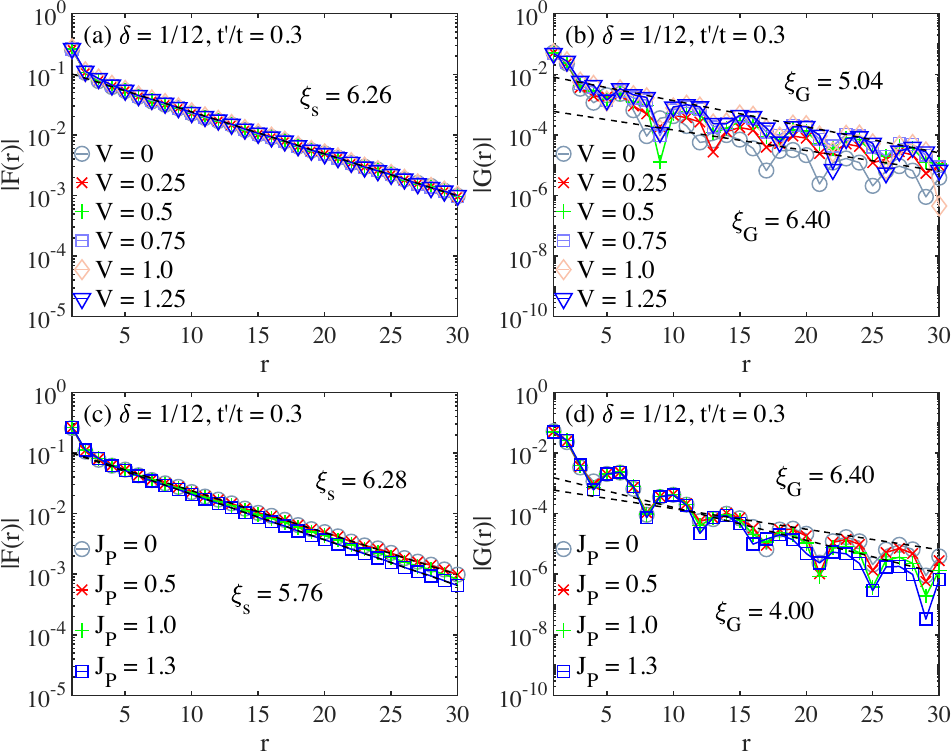}
   \caption{\label{SM_spin_green}
   Coupling dependence of the spin correlation $F(r)$ and single-particle Green's function $G(r)$ in the SC + CDW phase. (a) and (b) show the results of $|F(r)|$ and $|G(r)|$, respectively, with increased $V$ coupling for $\delta=1/12$, $t'/t = 0.3$. The correlation length $\xi_{\mathrm{s}} = 6.26$ in (a) is extracted by exponential fitting of $F(r)$ at $V=1.25$. $\xi_{\mathrm{G}}$ in (b) are extracted by exponential fitting of $G(r)$ at $V=0$ and $V=1.25$. (c) and (d) show the similar results with increased $J_P$ coupling. $\xi_{\mathrm{s}}$ in (c) are extracted by exponential fitting of $F(r)$ at $J_{P}=0$ and $J_{P}=1.3$. $\xi_{\mathrm{G}}$ in (d) are extracted by exponential fitting of $G(r)$ at $J_{P}=0$ and $J_{P}=1.3$. All these measurements are obtained from $L_x =48$ and by keeping bond dimensions $D=6000$.
}
\end{figure}

In Fig.~\ref{SM_spin_green}, we further show the spin correlation $F(r)$ and single-particle Green’s function $G(r)$ in the SC + CDW phase. 
For increased $V$ coupling, the fitted correlation lengths for both $F(r)$ and $G(r)$ appear to be insensitive to the coupling. 
On the other hand, for increased $J_P$, the fitted correlation lengths decrease slightly.

\section{\label{sec:Tuning_spin_exchange} Tuning spin exchange and density interactions in the $t$-$t'$-$J$ model}

\begin{figure}[t]
	\includegraphics[width=1.0\linewidth]{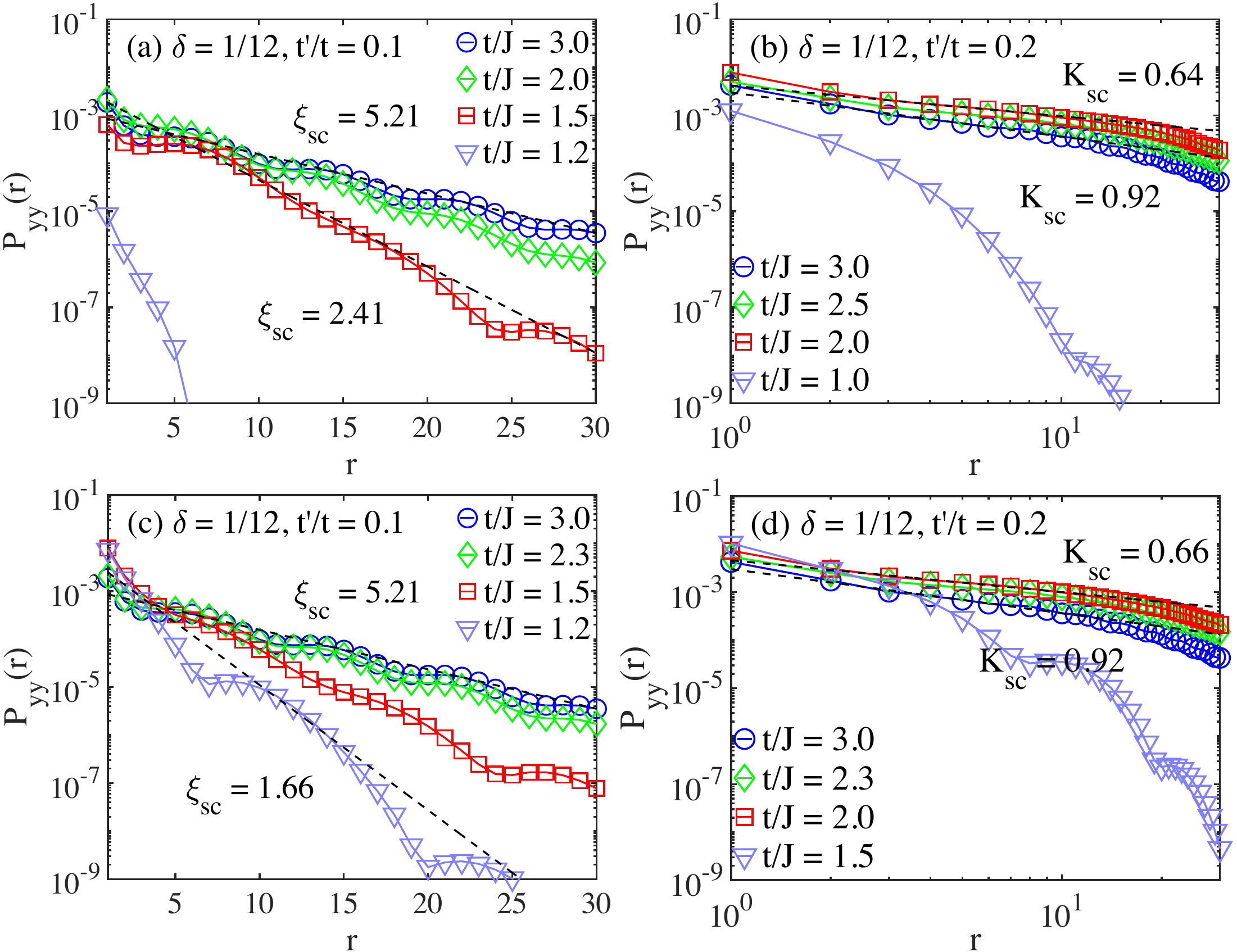}
	\caption{\label{tJ_SC}SC pairing correlation $P_{yy}(r)$. (a) and (b) are respectively the semi- and double-logarithmic plot of $P_{yy}(r)$ with different $t/J$ in the $t$-$t'$-$J$ model. (c) and (d) are the same plots of $P_{yy}(r)$ with different $t/J$ by fixing the charge density attraction strength ($\frac{1}{4}\sum_{\left\langle ij\right\rangle } {\hat{n} }_i {\hat{n} }_j$). The correlation length $\xi_{\mathrm{sc}}$ and power exponent $K_{\mathrm{sc}}$ are respectively obtained by exponential fitting and algebraic fitting. Here, all the data are obtained from $L_x =48$ and by keeping bond dimensions $D=6000$.}
\end{figure}

\begin{figure}[t]
	\includegraphics[width=1.0\linewidth]{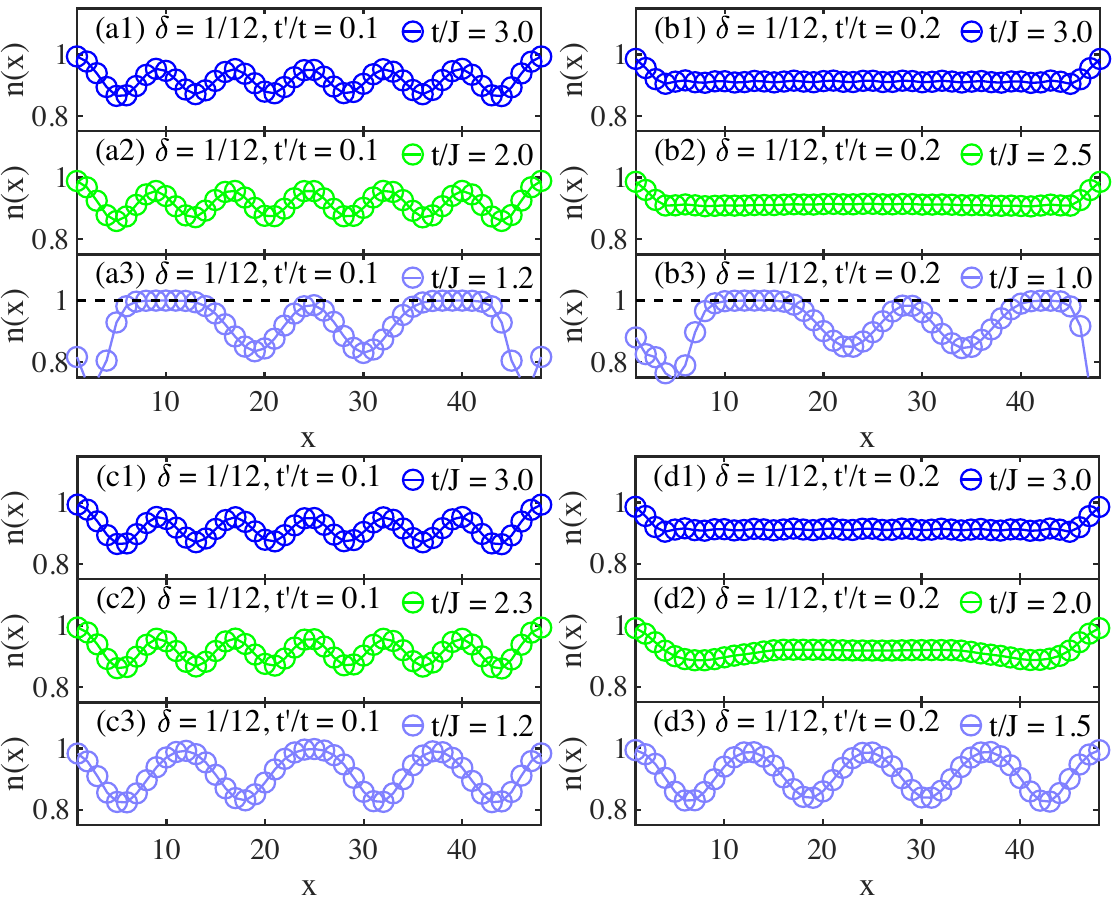}
	\caption{\label{tJ_CDW}Charge density profile $n(x)$. (a) and (b) are the $n(x)$ with different $t/J$ in the $t$-$t'$-$J$ model. (c) and (d) are the same plots of $n(x)$ with different $t/J$ by fixing the charge density attraction strength ($\frac{1}{4}\sum_{\left\langle ij\right\rangle } {\hat{n} }_i {\hat{n} }_j$). The black dash lines in (a3) and (b3) denote $n(x)=1$. Here, all the data are obtained from $L_x =48$ and by keeping bond dimensions $D=6000$.}
\end{figure}

In the $H_{J_P}$ Hamiltonian Eq.~\eqref{HtJ2}, $J_P$ has contributions to both spin exchange and density interactions.
Since the contributed repulsive density interaction $J_P \hat{n}_i \hat{n}_j / 4$ usually tends to result in charge orders~\cite{PS_ZZ,CDW_V_Santos,CDW_V_Sandvik,CDW_V_Greco,CDW_V_Alexandre,Huang_PRB_2013,Clay_CDW_2023,Abram_JPCM_2017}, here we further examine the $t$-$t'$-$J$ model by increasing only $J$ coupling.
We start from the system with $\delta =1/12$, $t/J = 3.0$ and $t'/t = 0.1$, which sits in the partially filled stripe phase but near the phase boundary with the SC phase [see Fig.~\ref{Pha_Dia}].
We find that pairing correlations decay faster with increased $J$ [Fig.~\ref{tJ_SC}(a)], and a larger $J$ term will lead to a phase separation as shown by the charge density profile in Fig.~\ref{tJ_CDW}(a1-a3), which agrees with the previous result of the $t$-$J$ model~\cite{PS_Lin,Dagotto_PRL_1993}.
For a comparison, we have also examined the response of the system in the SC phase (with $t'/t = 0.2$).
Similar to Fig.~\ref{SC_Pairing_Correlation}(c-d), pairing correlation is enhanced before the system transits to phase separation [see Fig.~\ref{tJ_SC}(b) and Fig.~\ref{tJ_CDW}(b1-b3)].

Furthermore, to clarify the consequences from either the Heisenberg term $J \hat{{\bf S}}_i \cdot \hat{{\bf S}}_j$ or density attraction $-J {\hat{n} }_i {\hat{n} }_j / 4$, we fix the density interaction and only increase the Heisenberg term. 
It turns out that we find the similar results as tuning the Heisenberg and density interactions simultaneously [Fig.~\ref{tJ_SC}(c-d)].
The only difference is that in this case the system will enter the filled stripe phase [Fig.~\ref{tJ_CDW}(c1-c3) and Fig.~\ref{tJ_CDW}(d1-d3)] instead of phase separation, when the partially filled stripe phase disappears.


\section{\label{sec:Hole_doped} DMRG results at $t'<0$}

\begin{figure}[h]
   \includegraphics[width=1.0\linewidth,angle=0]{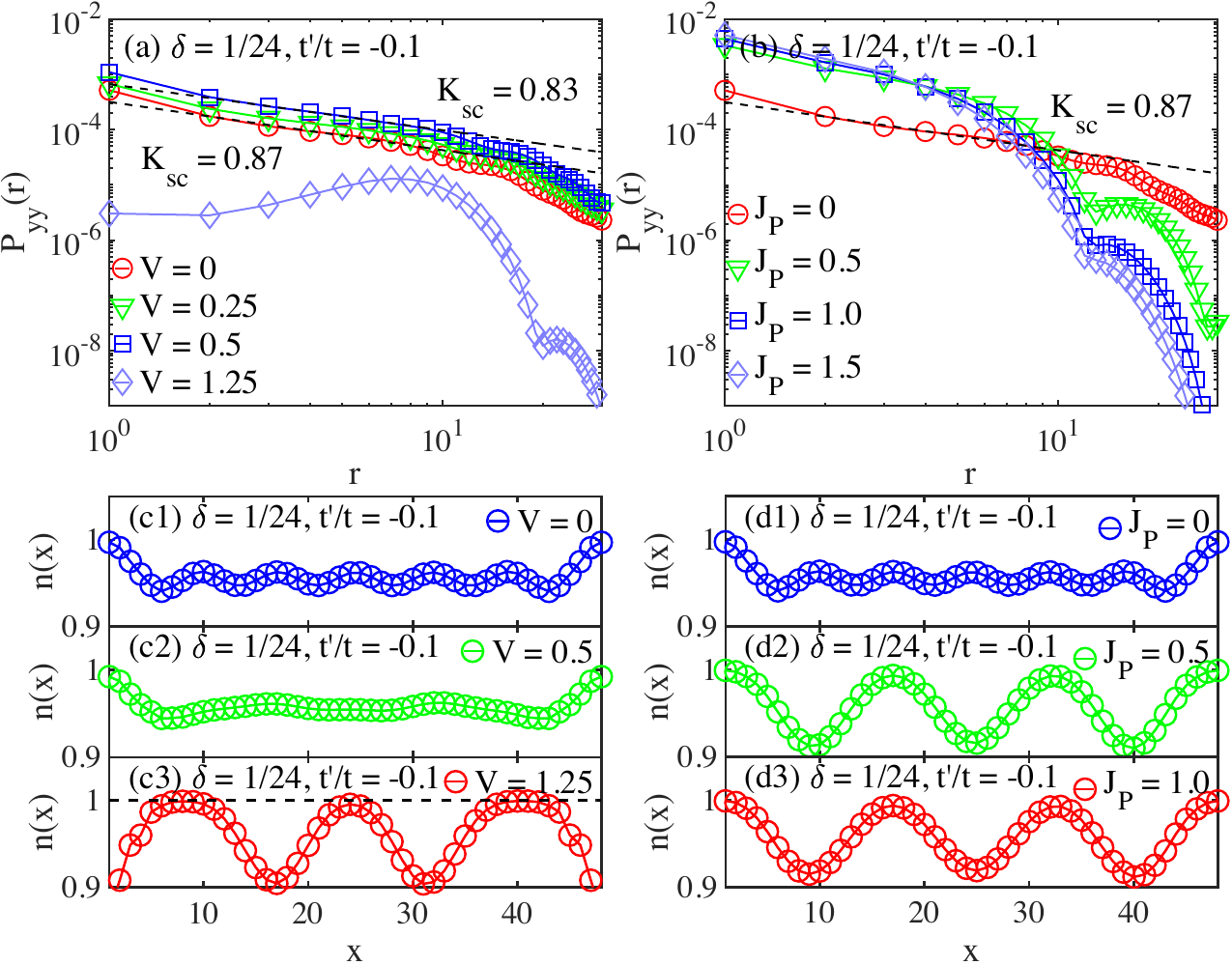}
   \caption{\label{SM_hole_doped}
   Pairing correlation and charge density profile of the system with $\delta = 1/24$, $t'/t = -0.1$ on the six-leg cylinder. (a) and (b) show the pairing correlation $P_{yy}(r)$ with increased $V$ and $J_P$ couplings, respectively. The power exponents $K_{\mathrm{sc}}$ in (a) are extracted by algebraic fitting of $P_{yy}(r)$ at $V=0$, $V=0.5$. $K_{\mathrm{sc}}$ in (b) is extracted by algebraic fitting of $P_{yy}(r)$ at $J_{P}=0$. (c) and (d) show the charge density profile $n(x)$ with increased couplings. All these measurements are obtained from $L_x =48$ and by keeping bond dimensions $D=6000$.
}
\end{figure}

For $t' < 0$ on the six-leg cylinder system, recent DMRG work also found a SC + CDW phase near the doping level $\delta = 1/24$~\cite{lu2023emergent}. 
We also examine the response of this SC + CDW phase to the increased $V$ or $J_P$ coupling.
We select a parameter point $t'/t = -0.1$, $\delta = 1/24$ to demonstrate our DMRG results. 
In Fig.~\ref{SM_hole_doped}(a), we show that the increased $V$ coupling could enhance the pairing correlation, and the CDW profile is suppressed [Fig.~\ref{SM_hole_doped}(c2)], which appears similar to the response of the SC + CDW phase at $t' > 0$. 
When the attraction $V$ is strong enough, the system also has a transition to the phase separation [Fig.~\ref{SM_hole_doped}(c3)]. 
With increased $J_P$, we find that a small $J_P$ can drive the system to the partially filled stripe phase [Fig.~\ref{SM_hole_doped}(d)] with exponentially decaying pairing correlation [Fig.~\ref{SM_hole_doped}(b)].

We have also checked the partially filled stripe phase at $t' < 0$ on the six-leg system.
With increased $V$ or $J_P$ coupling, we do not find SC phase either.

\clearpage

\twocolumngrid
\bibliography{refs}

\begin{thebibliography}{128}
\bibitem [{not()}]{note}%
  \BibitemOpen
  \href@noop {} {\bibinfo {title} {We have also examined the {SSH-Hubbard}
  model in the antiadiabatic limit on the four-leg cylinder. {F}or the fixed
  $\mathit{U}/t = 8$ at $1/8$ filling, we only find a stable stripe phase with
  increasing $\mathit{J}/t$, which is consistent with our results at the
  large-$\mathit{U}$ limit.}}\BibitemShut {Stop}%
\end{thebibliography}

\end{document}